\documentclass{article}
\usepackage{amsmath}
\begin{document}

\title{Nonrelativistic and Relativistic Continuum Mechanics}

\author{\it R. Beig\\
Institute for Theoretical Physics,
University of Vienna
\\ Vienna, Austria\\
E-mail: robert.beig@univie.ac.at}

\maketitle

\abstract{%
There is described a spacetime formulation of both nonrelativistic
and relativistic elasticity. Specific attention is devoted to the
causal structure of the theories and the availability of local
existence theorems for the initial-value problem. Much of the
presented material is based on joint work of B.G.Schmidt and the
author (in Class.Quantum Grav.\textbf{20} (2003), 889-904).

\section{Introduction}  \label{intro}

The ancient field theory of continuum mechanics, created by the
mathematicians J.Bernoulli, Euler and Cauchy, has grown into a
subject of great importance both for its mathematical interest and
its applications in material science. In its relativistic guise
this theory has not been developed very far yet. Important
references are \cite{SOU}, \cite{CQ}, \cite{C}, \cite{KM},
\cite{T} and \cite{KA}. The book \cite{SOP} is also an excellent
source.

We start, in Section \ref{nonr}, by describing the nonrelativistic
theory in a framework akin to that used in relativity, namely that
of Galilean spacetimes. In the following Section  \ref{hyp} we
describe the concept of  hyperbolicity appropriate for the
resulting system of 2nd-order partial differential equations. We
then describe a way of rewriting such a system in symmetric
first-order form. The condition of symmetric hyperbolicity for
this latter system, although true in many cases of physical
interest, is however a more stringent requirement than that of
hyperbolicity for the original second-order system.
 In Section \ref{nat} we write down a class of states,
the so-called "natural states", which satisfy the assumptions of
(both the first- and second-order) hyperbolic theory outlined in
Section \ref{hyp} so that there is available, for initial states
sufficiently close to natural ones, a local existence theorem for
the Cauchy problem.\footnote{There are in fact theorems even on
global existence (see \cite{S})} Finally, in Section \ref{rel}, we
describe the necessary changes as one goes from the
nonrelativistic theory (or rather: "Galilean relativity") to
Einsteinian relativity.

\section{Nonrelativistic Theory} \label{nonr}

We start with a Galilean spacetime $M$(see e.g \cite{EHL}). This
is furnished by $\bf{R^4}$, endowed with a symmetric, degenerate
contravariant metric $h^{\mu \nu}$ of signature $(0+++)$ and a
choice of covector field $\tau_\mu$ satisfying

\begin{equation}
\label{time} h^{\mu\nu} \tau_{\nu} = 0
\end{equation}
We also assume we are given a flat connection $\nabla_{\mu}$ which
annhilates both $h^{\mu\nu}$ and $\tau_{\mu}$. The matter flow is
described by a vector field $v^{\mu}$ normalized by $v^{\mu}
\tau_{\mu} = 1$, i.e. of the form
\begin{equation}
\label{vel}
v^{\mu}\partial_{\mu} = \partial_t + v^i \partial_i,
\end{equation}
where $v^i = v^i (t,x^j)$ and $(x^{\mu})=(t,x^i)$ are flat
coordinates in which $h^{\mu \nu}\partial_{\mu}\partial_{\nu} =
\delta^{ij}\partial_i\partial_j$ and $\tau_{\mu}dx^{\mu}= dt$. The
given flat connection $\nabla$ is not the only one annihilating
$(h^{\mu\nu},\tau_{\rho})$. One can use this freedom to describe
the effect of gravity by using the  connection  $\bar \nabla$,
where $\bar \nabla_{\mu} \omega_{\nu} = \nabla_{\mu} \omega_{\nu}
- C^{\lambda}_{\mu\nu} \omega_{\lambda}$ with
$C^{\lambda}_{\mu\nu} = \tau_{\mu}\tau_{\nu}h^{\lambda \sigma}
\nabla_{\sigma} U$, $U$ being the gravitational potential. One
easily checks that the connection $\bar \nabla$ again annihilates
$(h^{\mu\nu},\tau_{\nu})$. Having described the kinematical arena,
we now turn to the specific class of physical models we consider.
We start by writing down a ``stress-mass'' tensor. This tensor is
not such a natural object in the nonrelativistic theory as the
stress-energy tensor is in relativity, for two reasons: the lack
of a non-degenerate spacetime metric and the fact that the
Lagrangian of the nonrelativistic theory breaks the Galilean
invariance. We will nonetheless use the concept of stress-mass
here, since it greatly facilitates the task of moving back and
forth between the Galilean and the Einsteinian theory.

The mass-stress tensor has the form
\begin{equation}
\label{stress}
T^{\mu\nu} = \rho v^{\mu}v^{\nu} + t^{\mu\nu},
\end{equation}
where the Cauchy stress tensor $t^{\mu\nu}$ is purely spatial in
the sense that $t^{\mu\nu}\tau_{\nu} = 0$. Furthermore we have
that $\rho = n m_0$, where $n > 0$ is the particle number density
and $m_0$ the mass per particle. The continuity equation in our
language takes the following form. Take $\varepsilon$, the volume
form on $M$ defined in the adapted coordinates as
$\varepsilon_{ijk0}=\epsilon_{ijk}$, and consider the three-form
$N$ given by $N_{\mu \nu \lambda} = n \varepsilon_{\mu \nu \lambda
\rho} v^\rho$. Then conservation of mass is given simply by $d N =
0$. In the standard coordinates, $N$ is given by
\begin{equation}
\label{volume}
N = n \epsilon_{ijk}(dx^i - v^i dt)(dx^j - v^j dt)(dx^k - v^k dt)
\end{equation}
and $d N = 0$
 is of course equivalent to
\begin{equation}
\label{cont}
\partial_{\mu}(n v^{\mu}) = \partial_t n + \partial_i (n v^i) = 0.
\end{equation}
The matter field equations are given by
\begin{equation}
\label{field}
\nabla_{\nu} T^{\mu\nu} = \partial_{\nu}T^{\mu\nu} = 0
\end{equation}
in the absence of gravity, otherwise the derivative
$\bar{\nabla}_{\mu}$ has to be used in Eq.(\ref{field}). We
immediately see that the equation
$\tau_{\mu}\nabla_{\nu}T^{\mu\nu}=0$ is already implied by the
continuity law Eq.(\ref{cont}). In order to turn Eq.(\ref{field})
into proper field equations we have to specify the dependent
variables. These are furnished by maps $f$ sending points of
spacetime $M$ into a manifold $\bf{B}$ called body or material
manifold. This material manifold should be viewed as an abstract
set of labels which parametrize the particles making up the
continuum.  Thus $f$ is the ``back-to-labels-map''. If we choose
coordinates $(X^A)$ with $A=1,2,3$ on $\bf{B}$, we can write $X^A
= f^A(t,x^i)$. The relationship between the map $f^A$ and the
vector field $v^{\mu}$ is given by
\begin{equation}
\label{vel1} v^{\mu} \partial_{\mu}f^A = \partial_t f^A + v^i
\partial_i f^A = 0
\end{equation}

Suppose that $\bf{B}$ is endowed with a volume form $\Omega_{ABC}$
and set
\begin{equation}
\label{n} f^* \Omega = N,
\end{equation}
where $f^*$ denotes pull back under the map $f$. Eq.(\ref{n})
defines $n$ in terms of $f^A$, namely there holds $n =
det(\partial_i f^A)$. We assume the map $f^A$ to be such that
$f(t,.)$ is a diffeomorphism onto its image in $\bf{B}$ for all
$t$ and oriented so that $n$ is positive. This implies that the
map $f$ is of maximal rank. Hence, given $f$, Eq.(\ref{vel1}) has
a unique solution $v^i$. Here, and in what follows, the spacetime
field $v^{\mu}$ is always viewed as a function of
$(f^A,\partial_\mu f^B)$.  We now introduce the concept of
``strain'' by means of quantities $H^{AB}$ defined by
\begin{equation}
\label{strain}
H^{AB} = h^{\mu\nu}(\partial_{\mu}f^A)(\partial_{\nu}f^B)
\end{equation}
Clearly $H^{AB}$ is positive definite. Consequently there exists the inverse
$H_{AB}$ defined by
\begin{equation}
\label{invstrain}
H^{AB}H_{BC}=\delta^A{}_C
\end{equation}
We now assume that the Cauchy stress tensor in Eq.(\ref{stress}) is of the form
\begin{equation}
\label{cauchy} t^{\mu\nu}= n \tau_{AB}(\partial_{\rho}f^A)
(\partial_{\sigma}f^B) h^{\rho\mu} h^{\sigma\nu}
\end{equation}
with
\begin{equation}
\label{tau} \tau_{AB}= 2 \frac{\partial e}{\partial H^{AB}}
\end{equation}
for some function $e = e(f^A(x), H^{BC}(x))$, called
stored-energy function in the elastic literature. As an example
take the case where $e$ just depends on $n$. Note this makes sense
since $n$ can be written in terms of $H^{AB}$, namely
\begin{equation}
\label{nH}
6 n^2 = H^{AA'}H^{BB'}H^{CC'}\Omega_{ABC}\Omega_{A'B'C'}
\end{equation}
When $e$ depends only on $n$ one finds that
\begin{equation}
\label{fluid}
t^{\mu\nu} = p h^{\mu\nu},
\end{equation}
where $p$ is defined by
\begin{equation}
\label{press}
p=n^2 \frac{\partial e}{\partial n}
\end{equation}
The field equations here are the Euler equations for a perfect fluid.

For completeness we outline the proof of Eq.(\ref{fluid}). We
first claim that
\begin{equation}
\label{NH} \frac{\partial n}{\partial H^{AB}} = \frac{n}{2} H_{AB}
\end{equation}
The trick how to obtain Eq.(\ref{NH}) is to first compute
$\frac{\partial n}{\partial H^{AB}} H^{BC}$, using Eq.(\ref{nH}).
One quickly finds that
\begin{equation}
\label{NH1} \frac{\partial n}{\partial H^{AB}} \ H^{BC} =
\frac{n}{2} \ \delta_A{}^C
\end{equation}
We next define a quantity $F^\mu{}_A$ by
\begin{equation}
\label{Fmu} F^\mu{}_A = (\partial_\nu f^B)h^{\mu \nu}H_{AB}
\end{equation}
One checks that
\begin{equation}
\label{finv} (\partial_\mu f^A) F^\mu{}_B = \delta ^A{}_B
\end{equation}
Using $F^\mu{}_A \tau_\mu =0$ and $(\partial_\mu f^A) v^\mu =0$,
it follows that
\begin{equation}
\label{finv1} F^\mu{}_A (\partial_\nu f^A) = \delta^\mu{}_\nu -
v^\mu \tau_\nu
\end{equation}
Using Eq.'s (\ref{NH},\ref{finv1}) in Eq.(\ref{cauchy}), we
immediately obtain Eq.(\ref{fluid}) together with (\ref{press}).

 Let us return to the case of a general elastic
solid. The field equations are of the following form
\begin{equation}
\label{field1}
 \frac{\partial T^{\mu \lambda}}{\partial(\partial_{\nu} f^A)} \ \partial_{\lambda}
\partial_{\nu}f^A =
\textrm{lower-order derivatives of}  \ f^A,
\end{equation}
which, by the remark following Eq.(\ref{field}), are equivalent to
\begin{equation}
\label{field3} M^{\mu \nu}_{AB}\ \partial_{\mu} \partial_{\nu} f^B
= \mathcal{G}_A
\end{equation}
where $M^{\mu \nu}_{AB}$, defined as
\begin{equation}
\label{symbol} M^{\mu \nu}_{AB} = (\partial_{\lambda}f^C)  H_{CA}
\frac{\partial T^{\mu \lambda}} {\partial(\partial_{\nu} f^B)},
\end{equation}
and $\mathcal{G}_A$ are functions of $(f^A,\partial_{\mu}f^B)$.
Remarkably the quantities $M^{\mu \nu}_{AB}$ turn out to satisfy
the symmetry
\begin{equation}
\label{sym}
M^{\mu \nu}_{AB} = M^{\nu \mu}_{BA}
\end{equation}
The symmetry Eq.(\ref{sym}) is no accident. It is due to the fact
that the field equations are the Euler-Lagrange equations of a
variational principle. Explicitly we find
\begin{equation}
\label{M} M^{\mu\nu}_{AB}= -\rho v^{\mu}v^{\nu} H_{AB} + n
[\tau_{AB}H_{CD}+2 \tau_{C(A}H_{B)D}+2\frac{\partial
\tau_{AC}}{\partial
H^{BD}}]\partial^{\mu}\!f^C\partial^{\nu}\!f^D,
\end{equation}
where we have defined
$\partial^{\mu}\!f^A=h^{\mu\nu}\partial_{\nu}f^A$. The associated
Lagrangian, which in particular satisfies
\begin{equation}
\label{EL} - M^{\mu \nu}_{AB} = \frac{\partial^2
\mathcal{L}}{\partial (\partial_{\nu}f^B)
\partial (\partial_{\mu}f^A)} \ ,
\end{equation}
is given by
\begin{equation*}
\label{L} \mathcal{L} = n ( \frac{1}{2} m_0 v^i v^j \delta_{ij} -
e ) .
\end{equation*}
Note that the quantities $n,e,v^i$ should be all regarded as
functions of $(f^A, \partial_{\mu}f^B)$.

This ends our description of nonrelativistic elasticity in its
spatial (or rather:"spacetime") form. We remark that all standard
treatments in the literature (see e.g. \cite{G} or \cite{MH})
prefer the material form based on the map $F^i (t,X^A)$ defined by

\begin{equation}
\label{spatial} f^A(t,F^i(t,X^B)) = X^A.
\end{equation}

From the relativistic point of view the spacetime form is
preferable, since, in a relativistic spacetime, there does not
exist the standard $t=const$-foliation available in Galilean
spacetime.

\section{Some Hyperbolic Theory}  \label{hyp}
An equation of the form of (\ref{field3}) will be called
hyperbolic if $M^{\mu \nu}_{AB}$ satisfies the symmetry
(\ref{sym}) and the following holds: firstly there should exist a
subcharacteristic covector, i.e. a covector $\xi_{\mu}$ so that
\begin{equation}
\label{subch}
M^{\mu \nu}_{AB} \xi_{\mu}\xi_{\nu} \textrm{ is negative definite}
\end{equation}
Secondly there should exist a timelike vector, i.e. a vector $X^{\mu}$
so that
\begin{equation}
\label{timelike} M^{\mu \nu}_{AB} \eta_{\mu}m^A \eta_{\nu}m^B
\textrm{ is positive definite}
\end{equation}
for all nonzero $\eta_{\mu}$ with $X^{\mu}\eta_{\mu}=0$. Let us
pause for a moment to explain by means of an example simpler than
elasticity how the change of sign between Eq.(\ref{subch}) and
Eq.(\ref{timelike}) arises. The example is the equation for wave
maps, where $M^{\mu \nu}_{AB}=g^{\mu \nu} G_{AB}$ with $g^{\mu
\nu}$ the spacetime metric and $G_{AB}$ the Riemannian metric on
the target space. Now the notion of timelike has its standard
Lorentzian meaning, and subcharacteristic covectors are timelike
covectors. The sign change between (\ref{subch}) and
(\ref{timelike}) is then simply due to the fact that (co-)vectors
orthogonal to a timelike (co-)vector are spacelike.

A vector $X^{\mu}$ is called causal if $X^{\mu} \xi_{\mu} \neq O$
for all subcharacteristic covectors $\xi_{\mu}$. Clearly all
timelike vectors are causal. In general there will be a gap
between timelike and noncausal vectors due for example to the
existence of different characteristic (sound) cones. A covector
$k_{\mu}$ is called characteristic if the symbol of the PD
operator on the l.h. side of Eq.(\ref{field3}), namely the
quadratic form $M_{AB}(k)$ given by
\begin{equation}
\label{charc}
M_{AB}(k) = M^{\mu\nu}_{AB}k_{\mu}k_{\nu}
\end{equation}
is degenerate. A vector $X^{\mu}$ is called (bi-)characteristic if
it is of the form
\begin{equation}
\label{charv}
X^{\mu} = M^{\mu\nu}_{AB}k_{\nu}m^A m^B
\end{equation}
for a characteristic covector $k_{\mu}$ and $m^A$ such that
$M^{\mu\nu}_{AB}k_{\nu}k_{\nu} m^B =0$. This characteristic vector
is tangent to the characteristic sheet to which $k$ belongs where
this sheet is a regular surface.

The above definitions are essentially taken from the book
\cite{CHR}.\footnote{We have only added the word "timelike" for
vectors having the property in Eq.(\ref{timelike}) and
"subcharacteristic" for covectors satisfying Eq.(\ref{subch}).}
The notion of characteristic vector from \cite{CHR} as above is
new, the classical one breaking down when the set of
characteristic covectors (also called "normal cone" or "slowness
cone" in the literature) has singularities due to intersections
between different sheets (there are in general three such sheets
which are given by the zero-level set of the eigenvalues of
$M_{AB}$). Even in the absence of singularities of the normal cone
the set of characteristic vectors (also called "ray cone" or "wave
cone" in the literature) will in general have cusps (see e.g.
Chapter VI of \cite{CH}).

A yet more general notion of hyperbolicity, due to Kreiss, of
which the above is a special case, is that of strong hyperbolicity
(see \cite{KR}).

A key point regarding these definitions is the availability of an
existence theorem independently of the form of the lower-order
terms in Eq.(\ref{field3}).\footnote{In \cite{CH} a result to that
extent is stated without proof.} Namely, suppose there is a
hypersurface $S$ of spacetime, together with initial data for
$f^A$ and $\partial_{\mu}f^A$ on $S$ so that, for these data, the
surface $S$ has everywhere subcharacteristic conormal. Then choose
a vector field $X^{\mu}$ which is timelike on $S$, whence
transversal to $S$. Use this vector field to Lie-drag $S$ into the
future. Since the properties of being subcharacteristic and
timelike are
 "open" conditions, we thus obtain a spacetime
neighbourhood $N\subset M$ of $S$, which is foliated by surfaces
which are subcharacteristic for all maps $f$ close to the initial
one and where $X^{\mu}$ is timelike with respect to such
configurations. Now take coordinates $(y^0,y^i)$ so that the
leaves of the foliations are given by $y^0=const$ and the vector
field $X^{\mu}$ is given by $X^{\mu}\partial_{\mu} =
\partial_0$. In these coordinates the equation (\ref{field3}) takes the
form
\begin{equation}
\label{kato} [M^{00}_{AB} \partial^2_0 +
(M^{0i}_{AB}+M^{i0}_{AB})\partial_0\partial_i+M^{ij}_{AB}\partial_i\partial_j]f^B
= \textrm {lower-order derivatives of} f^A.
\end{equation}
Furthermore there should hold
\begin{equation}
\label{kato1} M^{00}_{AB}m^A m^B<0,\ M^{ij}_{AB}l_i l_j m^Am^B>0
\end{equation}
with $l_i, m^A$ both nonzero and all values of
$(f^A,\partial_{\mu}f^B)$ close to those corresponding to the
initial data. If the neighbourhood $N$ is of the form $\{y^0 \in
[0,T]\} \times \{y \in \mathbf R^3 \}$ and the initial data
satisfy some decay properties for large $\arrowvert y \arrowvert$
one can now appeal to a basic theorem in \cite{HU} to infer
existence of a unique solution for sufficiently small $T$. (We do
not spell out the precise differentiability requirements.) The
asymptotic conditions imposed in the above theorem are of course
not always appropriate, and one would in any case like a local
statement amounting to uniqueness in the "domain of dependence" of
initial data in  open subsets of $S$. Such a theorem is proved in
\cite{CHR}, the appropriate notion of domain of dependence being
based on causal curves (in the sense of causal vectors as
described above). Consequently the nonlocal nature of the
uniqueness part of the theorem in \cite{HU} is in fact irrelevant.

In \cite{BS} the autors chose to cast the equations of elasticity
theory into that of a first-order symmetric hyperbolic system,
which goes as follows: Define 5-index quantities
$W^{\mu\nu}{}_{AB}{}^{(\lambda)}$ by

\begin{equation}
\label{W} W^{\mu\nu}{}_{AB}{}^{(\lambda)}:= X^\mu
M^{\lambda\nu}{}_{AB}-2X^{[\lambda}M^{\nu]\mu}{}_{BA},
\end{equation}
where $X$ is a timelike vector. We now replace (\ref{field3}) by
the following first-- order system:
\begin{equation}
\label{first}
 W^{\mu\nu}{}_{AB}{}^{(\lambda)}(f,F)\partial_\lambda
F^B{}_\nu  = X^\mu(f,F) \mathcal{G}_A(f,F)
\end{equation}
\begin{equation}
\label{first1}
 -X^{\lambda}(f,F)\partial_{\lambda}f^A =
-X^{\lambda}(f,F)F^A{}_{\lambda}
\end{equation}
together with the constraint $\partial_{\mu}f^A=F^A{}_{\mu}$.
Since
\begin{equation}
\label{W1}
W^{\mu\nu}{}_{AB}{}^{(\lambda)}=W^{\nu\mu}{}_{BA}{}^{(\lambda)},
\end{equation}
the system (\ref{first},\ref{first1}) is symmetric. One then finds
that (\ref{first},\ref{first1}) is equivalent to the original
second-order system (\ref{field3}).\footnote{In (\cite{BS} we used
a special property of elasticity to prove this equivalence. It is
not hard to see that this equivalence works generally for the
system (\ref{field3}) when $X^\mu$ is timelike.} Next recall that
the symmetric system is called symmetric hyperbolic if there
exists a subcharacteristic covector $\xi_\mu$, i.e. one so that
the quadratic form defined by the left-hand side of the system
Eq.(\ref{first}), i.e.
$W^{\mu\nu}{}_{AB}{}^{(\lambda)}\xi_\lambda$ is negative definite
in the variables $m^A{}_\mu$. Suppose $\xi$ is subcharacteristic
for the second-order system and $X^\mu \xi_\mu > 0$: is it then
subcharacteristic also for the system (\ref{first},\ref{first1})?
The answer in general is "no" as we will see in the next section.

\section{Natural States}  \label{nat}

We now further specify the "equation of state" given by the
stored-energy function, as follows: We assume $\bf{B}$ to be
endowed with a flat Riemannian metric $G_{AB}$ and that the volume
form $\Omega$ is compatible with $G_{AB}$. The stored-energy
function $e$ is then assumed to be of the form $e=e(H^{AB})$,
where $H^{AB}$ now refers to coordinates $X^A$ on $\bf{B}$ in
which $G_{AB}=\delta_{AB}$. (Note this implies that the field
equations (\ref{field3}) and (\ref{first}) have
$\mathcal{G}_A=0$.) More specifically we suppose $e$ to satisfy
\begin{equation}
\label{constants}
 e= \frac{1}{8} E_{ABCD}(H^{AB}-\delta^{AB})(H^{CD}
- \delta^{CD})+ O((H-\delta)^3)
\end{equation}
for certain constants $E_{ABCD}=E_{(AB)(CD)}=E_{CDAB}$. Clearly
there are 21 independent such constants available. One furthermore
assumes these constants to be such that
\begin{equation}
\label{hadamard} E_{ABCD}l^A m^B l^C m^D > 0
\end{equation}
for $m^A, l^A$ both nonzero. The definition, then, of a natural
state $\stackrel{\circ}f$ is that of a map
$\stackrel{\circ}f{}\!\!^A$ which corresponds to a configuration
of zero strain in that
\begin{equation}
\label{iso}
\stackrel{\circ}H{}\!^{AB}(x)=(\partial_{\mu}\!\!\stackrel{\circ}f{}\!\!^A(x))
(\partial_{\nu}\!\!\stackrel{\circ}f{}\!\!^B(x)) h^{\mu\nu}(x)=
\delta^{AB}.
\end{equation}
Note this relation implies
\begin{equation}
\label{rigid} \mathcal{L}_{\stackrel{\circ}v}h^{\mu \nu} = 0
\end{equation}
i.e. that $\stackrel{\circ}v^{}\!^\mu$ is a rigid motion. The
absence of linear terms in Eq.(\ref{constants}) furthermore
implies that natural states are stressfree, i.e. have
$\tau_{AB}=0$. Using also that $\stackrel{\circ}n = 1$ we find
that
\begin{equation}
\label{Mnat} \stackrel{\circ}M{}\!^{\mu\nu}_{AB}= -m_0
\stackrel{\circ}v{}\!^\mu \stackrel{\circ}v{}\!^\nu \delta_{AB} +
E_{ACBD}R^{C\mu}R^{D\nu}
\end{equation}
with $R^{A\mu}=R^A{}_B \delta^{B\mu}$ and $R^A{}_B$ a (in general
time dependent) rotation matrix. The covector $\tau_\mu$, from
Eq.(\ref{Mnat}), is clearly subcharacteristic. Furthermore the
vector $\stackrel{\circ}v{}\!^\mu$ is timelike iff the conditions
(\ref{hadamard}) are valid. Therefore the second-order equations
are hyperbolic at natural states.

We now turn to hyperbolicity of the first-order system. Taking \\
$X^\mu = \stackrel{\circ}v{}\!^\mu$ and using Eq.s
(\ref{Mnat},\ref{W}), we see that
\begin{equation}
\label{Wnat} \stackrel{\circ}W{}\!^{\mu\nu}{}_{AB}{}^{(\lambda)}
\tau_\lambda = -m_0 \stackrel{\circ}v{}\!^\mu
\stackrel{\circ}v{}\!^\nu \delta_{AB} - E_{ACBD}R^{C\mu}R^{D\nu}
\end{equation}

Thus $\tau_\mu$ is subcharacteristic for natural states iff
\begin{equation}
\label{pos} E_{ABCD}m^{AB}m^{CD}> 0
\end{equation}
for all nonzero elements $m^{AB}=m^{(AB)}$, which is clearly a
stronger requirement than (\ref{hadamard}).

The only case which is easy to analyze fully is the isotropic case
\footnote{In \cite{BS} we mistakenly interchanged the constants
$\lambda$ and $\mu$ in equation (4.16) of that paper.} where
\begin{equation}
\label{isotropic} E_{ABCD}= \lambda \delta_{AB}\delta_{CD} + 2 \mu
\delta_{C(A}\delta_{B)D}
\end{equation}
The constants $\lambda,\mu$ in Eq.(\ref{isotropic}) are the
standard Lam\'e constants. They should not be confused with
indices $(\mu,\nu)$. The "rank-one convexity" condition
Eq.(\ref{hadamard}) is equivalent to
\begin{equation}
\label{rankone} m_0 c_2^2=\mu > 0, \ m_0c_1^2=2\mu + \lambda > 0
\end{equation}
The eigenvalues of $\frac{1}{m_0}M_{AB}(k)$ relative to
$\delta_{AB}$, which are real of course since $M_{AB}$ is
symmetric, are given by
\begin{equation}
\label{e1} \lambda_1(k) = c_1^2 \stackrel{1}g^{}\!^{\mu \nu} k_\mu
k_\nu
\end{equation}
and
\begin{equation}
\label{e2} \lambda_2(k) = \lambda_3(k) = c_2^2
\stackrel{2}g{}\!^{\mu \nu} k_\mu k_\nu,
\end{equation}
where
\begin{equation}
\label{ncone} \stackrel{1}g{}\!^{\mu \nu} = h^{\mu \nu} -
\frac{1}{c_{1}^2} \stackrel{\circ}v{}\!^\mu
\stackrel{\circ}v{}\!^\nu, \ \stackrel{2}g{}\!^{\mu \nu} = h^{\mu
\nu} - \frac{1}{c_{2}^2} \stackrel{\circ}v{}\!^\mu
\stackrel{\circ}v{}\!^\nu.
\end{equation}
(A Lorentzian metric of the above form is nowadays called
"acoustic metric" or "Unruh metric".) Thus the normal cone
consists of two sheets. The one corresponding to $\lambda_1$ is
associated with a longitudinal ("pressure") mode propagating at
speed $c_1$, the second one corresponding to two transversal
("shear") modes of speed $c_2$. If $c_2 < c_1$, the first sheet
lies inside the second sheet. Subcharacteristic covectors, for
which both $\lambda_1$ and $\lambda_2$ are negative, lie inside
the inner cone. One such covector is $\tau_\mu$. In fact,
$\tau_\mu$ lies on the central ray inside the two cones in the
following sense: the vector $v^\mu$ defines a family of parallel
hyperplanes in the cotangent space.These intersect the normal
cones in 2-surfaces which, in the metric $h^{\mu \nu}$, are
standard spheres centered at the point where the ray of $\tau_\mu$
intersects this hyperplane. The ray cones dual to the above normal
ones are given by the equations
\begin{equation}
\label{raycone} \stackrel{1}g_{\mu \nu}X^\mu X^\nu = (h_{\mu \nu}
- c_1^2 \tau_\mu \tau_\nu)X^\mu X^\nu, \ \stackrel{2}g_{\mu
\nu}X^\mu X^\nu = (h_{\mu \nu} - c_2^2 \tau_\mu \tau_\nu) X^\mu
X^\nu,
\end{equation}
where $h_{\mu \nu}$ is the unique tensor satisfying
\begin{equation}
\label{hinverse} h^{\mu \nu}h_{\nu \rho} = \delta^\mu {}_\rho -
v^\mu \tau_\rho, \ h_{\mu \nu}v^\nu = 0
\end{equation}
Note that $\stackrel{1}g_{\mu \nu}$ (resp. $\stackrel{2}g_{\mu
\nu}$) are the inverses of $\stackrel{1}g{}\!^{\mu \nu}$ (resp.
$\stackrel{2}g{}\!^{\mu \nu}$). Clearly, the cone of shear waves
is now the one lying inside. Timelike vectors X, for which all
covectors $k$ with $X^\mu k_\mu = 0$ have $\lambda_1$ and
$\lambda_2$ both positive, lie inside this inner ray cone. One
such timelike vector is $v^\mu$, in fact, it lies on the central
ray of the two ray cones in a fashion exactly dual to that
explained for $\tau_\mu$. Causal vectors may be "faster" in that
they lie inside or on the outer ray cone.

We now turn, finally in this section, to the question of
hyperbolicity of the first-order theory at natural isotropic
states. The condition (\ref{pos}) is valid if and only if
\begin{equation}
\label{possym} c_1^2 > \frac{4}{3} c_2^2 > 0,
\end{equation}
which is physically entirely reasonable, since elastic materials
typically have $c_1/c_2$ approximately 1,7. But one can do better
than that. One first notices that both the symmetries of $M^{\mu
\nu}_{AB}$ and the equation (\ref{field3}) remain untouched if the
quantities $M^{\mu \nu}_{AB}$ are replaced by $\bar{M}^{\mu
\nu}_{AB}$ given by
\begin{equation}
\label{Mbar} \bar{M}^{\mu \nu}_{AB} = M^{\mu \nu}_{AB} +
\Lambda^{\mu \nu}_{AB},
\end{equation}
where $\Lambda^{\mu \nu}_{AB} = \Lambda^{[\mu \nu]}_{AB} =
\Lambda^{\mu \nu}_{[AB]}$.(In fact this replacement can be viewed
as coming from adding a total divergence to the underlying
Lagrangian (see \cite{CHR})). While it is easily seen that
second-order hyperbolicity is unaffected by this replacement, this
is not the case for the associated first-order system. In the case
at hand we can, by adding to $E_{ACBD}$ a term of the form
\begin{equation}
\label{epsilon} (4c_2^2 - 2 \epsilon)\delta_{A[C}\delta_{D]B},
\end{equation}
arrange for Eq.(\ref{pos}) to hold if we take $\epsilon$ in the
range $0<\epsilon<min(2c_2^2,\frac{3c_1^2}{2})$, which of course
can alway be satisfied when $c_1$ and $c_2$ are both non-zero.

Let us point out that the standard equations of linearized
elasticity at a natural state are obtained by simply freezing the
coefficients in Eq.(\ref{field3}) and setting the right-hand side
equal to zero. The Cauchy problem for these equations is studied
e.g. in \cite{D}. In the isotropic case one finds that the
solution at $(t,x^i)$ does not depend on data at $t=0$ inside
$\vert x \vert < c_2 t$, i.e. inside the past inner ("shear") ray
cone.

\section{Relativistic Theory}  \label{rel}

Having available the spacetime form of the nonrelativistic theory
it is easy to write down its relativistic version, so we will be
brief, merely pointing out the necessary changes. We start out
with a relativistic spacetime $(M,g_{\mu \nu})$ with $g_{\mu \nu}$
a Lorentz metric. The configurations are again maps from spacetime
to $\bf{B}$, the latter endowed with a Riemannian metric $G_{AB}$
and compatible volume form $\Omega_{ABC}$. The maps $f$ should be
of maximal rank and such that the inverse image under $f$ of each
point of $\bf{B}$ in the image of $f$ is a timelike curve in $M$.
Thus there are timelike vectors $u^\mu$, unique up to scale, so
that
\begin{equation}
\label{u} (\partial_\mu f^A) u^\mu = 0
\end{equation}
We denote henceforth by $u^\mu$ the unique solution vector of
Eq.(\ref{u}) which is future-pointing and normalized by $g_{\mu
\nu}u^\mu u^\nu = -1$. The quantities $H^{AB}$ are defined as
\begin{equation}
\label{rstrain} H^{AB} = (\partial_\mu f^A)(\partial_\nu f^B)
g^{\mu \nu}.
\end{equation}
and are again positive definite. The particle number density $n$
is defined by
\begin{equation}
\label{n1} 6n^2 = \Omega_{A'B'C'} \Omega_{ABC}
H^{AA'}H^{BB'}H^{CC'}, \ n>0
\end{equation}
The Lagrangian of the theory (we set the speed of light equal to
one) is taken to be
\begin{equation}
\label{rL} \mathcal{L} = - \rho = -n(m_0 + e),
\end{equation}
with $e$ a function of $(f^A, H^{BC})$. Varying $\mathcal{L}$ with
respect to $g^{\mu \nu}$ gives the stress energy tensor
\begin{equation}
\label{stress1} T^{\mu \nu} = \rho u^\mu u^\nu +
2n\tau_{AB}(\partial^\mu f^A) (\partial^\nu f^B),
\end{equation}
where $\tau_{AB}= 2 \frac{\partial e}{\partial H^{AB}}$, as before
and $\partial^\mu f^A = g^{\mu \nu}\partial_\nu f^A$. The field
equations are again of the form Eq.(\ref{field3}) with
\begin{equation}
\label{M1} M^{\mu\nu}{}_{AB}=-\mu_{AB}u^\mu u^\nu
+U_{ACBD}F^{C\mu}F^{D\nu},
\end{equation}
where
\begin{equation}
\label{mu} \mu_{AB}= \rho H_{AB} + n \tau_{AB}
\end{equation}
and
\begin{equation}
\label{U}
U_{ACBD}=n(\tau_{AB}H_{CD}+\tau_{AC}H_{BD}+\tau_{BD}H_{AC}
+2\frac{\partial\tau_{BD}}{\partial H^{AC}})+2\rho H_{A[C}H_{D]B}.
\end{equation}
The last term in Eq.(\ref{U}) does not contribute to the equations
of motion. The second term on the right in Eq.(\ref{mu}) is
clearly a relativistic contribution.

There are slight complications in curved spacetime regarding the
notion of a natural state: in order for a natural state to exist,
the spacetime metric would have to allow Born rigid motions and
the material metric $G_{AB}$ would have to be isometric to the
metric on the quotient of $M$ by the action of this motion. We
avoid this difficulty here by confining ourselves to special
relativity  and taking the natural configuration to be at rest in
some inertial system. Thus we assume $(M,g_{\mu \nu})$ to be
Minkowski space. We also suppose the "natural" motion to be of the
form $\stackrel{\circ}u{}\!^\mu \partial_\mu =
\partial_t$ in coordinates $(t,x^i)$ in which $g_{\mu \nu} dx^\mu dx^\nu = -dt^2 + \delta_{ij}dx^i dx^j$.
We also assume  $G_{AB}=\delta_{AB}$ as before. Using (\ref{iso}),
a natural map corresponds to a time independent rotation in an
inertial system which, in the isotropic case, can be taken to be
the identity without loss. One then merely replaces, in the
expressions (\ref{e1},\ref{e2}) for the different cones, the
covector $\tau_\mu$ by the covector $-\stackrel{\circ}u_\mu =
-g_{\mu \nu}\stackrel{\circ}u{}\!^\mu$ and the symmetric tensor
$h^{\mu \nu}$ by $\bar{h}^{\mu \nu}$ given by
\begin{equation}
\label{barh}
 \bar{h}^{\mu \nu}=g^{\mu \nu} + \stackrel{\circ}u{}\!^\mu \stackrel{\circ}u{}\!^\nu.
\end{equation}
Furthermore one replaces the tensor $h_{\mu \nu}$ in
(\ref{hinverse}) by $\bar{h}_{\mu \nu}$ given by
\begin{equation}
\label{barh1} \bar{h}_{\mu \nu}= g_{\mu \nu} +
\stackrel{\circ}u_\mu \stackrel{\circ}u_\nu.
\end{equation}
Thus, writing $\bar{c}$ for either $c_1$ or $c_2$, the associated
normal cone is now given by the Lorentz metric
\begin{equation}
\label{normal} \bar{g}^{\mu \nu} = g^{\mu \nu} + (1 -
\frac{1}{\bar{c}^2})\stackrel{\circ}u{}\!^\mu
\stackrel{\circ}u{}\!^\nu
\end{equation}
and the ray cone by the inverse metric, namely
\begin{equation}
\label{ray} \bar{g}_{\mu \nu} = g_{\mu \nu} + (1 - \bar{c}^2)
\stackrel{\circ}u_\mu \stackrel{\circ}u_\nu.
\end{equation}
One easily infers from these relations that in the present
coordinates the special relativistic equations, linearized at a
natural state, are exactly identical with the nonrelativistic
ones, when the latter are written in coordinates where $v^\mu
\partial_\mu = \partial_t$ with $v^\mu$ an inertial motion. Of course the geometrical objects in both
theories are different - and thus the behaviour of the two
linearized theories under change of coordinates is also different.

{\bf{Acknowledgment}}: I thank H.-O.Kreiss for discussions and in
particular for motivating me to look more closely at the
second-order form of the elasticity equations. I also thank
B.G.Schmidt for helpful discussions, P.T.C. Chrus\'ciel for
comments and R.Low for suggesting some imporant improvements of
the manuscript.


\begin{thebibliography}{99}


\bibitem[1]{BS} {Beig R. and Schmidt B.G.} (2003) {\it Class. Quantum Grav.}
{\bf20} {889}

\bibitem[2]{CQ} {Carter B. and Quintana H.} (1972) {\it Proc.R.Soc.A}
{\bf331} {57}


\bibitem[3]{C} {Carter B.} (1973) {\it Commun.Math.Phys.}
{\bf30} {261}

\bibitem[4] {CH} {Courant R. and Hilbert D.} (1962) {\em Methods of Mathematical Physics,
Volume II\/}, {Interscience Publishers, New York}

\bibitem[5]{CHR} {Christodoulou D.} (2000) {\em The Action Principle and Partial
Differential Equations\/}, {Princeton University Press, Princeton NJ}

\bibitem[6] {D} {Duff G.F.G.} (1960) {\it Phil.Trans.R.Soc.Lond.A} {\bf252} {31}

\bibitem[7]{EHL} {Ehlers J.} (1998), in: {\em Understanding
Physics \/} {A.K.Richter (Ed.)}, {Copernicus Gesellschaft e.V.,
Katlenburg-Lindau}

\bibitem[8]{G} {Gurtin M.E.} (1981) {\em An Introduction to Continuum
Mechanics\/}, {Academic Press, New York}

\bibitem[9]{HU} {Hughes T.J.R., Kato T. and Marsden J.E.} (1977)
{\it Arch. Ration. Mech. Anal.} {\bf63} {273}

\bibitem[10] {KA} {Karlovini M. and Samuelsson L.} (2003) {\it
Class. Quantum Grav.} {\bf20} {3613}

\bibitem[11]{KM} {Kijowski J. and Magli G.} (1992) {\it
J. Geom. Phys.} {\bf9} {207}

\bibitem[12] {KR} {Kreiss H.-O.} (2002), in: {\em The Conformal
Structure of Spacetime\/}, {J.Frauendiener, H.Friedrich (Eds.)},
{Springer Lecture Notes in Physics, Berlin}

\bibitem[13] {MH} {Marsden J.E. and Hughes T.J.R.} (1994) {\em
Mathematical Foundations of Elasticity\/}, {Dover, New York}

\bibitem[14] {S} {Sideris T.C.} (2000) {\it Ann. of Math.(2)} {\bf151} {849}

\bibitem[15]{SOP} {Soper D.E.} (1976) {\em Classical Field Theory\/},
{Wiley, New York}

\bibitem[16]{SOU} {Souriau J.M.} (1958) {\em Publs. Scient. Univ. Alger.}
{\bf2} {103}

\bibitem[17] {T}  {Tahvildar-Zadeh A.S.} (1998) {\it Ann. Inst. H.
Poincar\'e} {\bf69} {275}

\end{thebibliography}
\end{document}